\documentclass[14.5pt,a4paper,UKenglish,cleveref, autoref, thm-restate]{article}
\usepackage[linesnumbered,ruled,vlined,titlenotnumbered]{algorithm2e}
\usepackage{graphicx}
 \usepackage{amsmath}
 \usepackage{fancyhdr}
 \usepackage{ amssymb }
 \usepackage{float} 
\graphicspath{ {./images/} }
\newcommand*{\field}[1]{\mathbb{#1}}%

\providecommand{\keywords}[1]{\textbf{\textit{Keywords---}} #1}

\title{On Unimodality of Independence Polynomials of Trees}

\author{
        Ron Yosef \\
                Department of Computer Science\\
        The Hebrew University of Jerusalem\\
         Jerusalem 9190401, \underline{Israel}\\
                  ron.yosef@mail.huji.ac.il\\

            \and
        Matan Mizrachi \\
                Department of Computer Science\\
        HIT - Holon Institute of Technology\\
         Holon, Golomb 52, \underline{Israel}\\
                 matanmi@my.hit.ac.il\\

            \and
        Ohr Kadrawi \\
                Department of Computer Science\\
        Ariel University\\
         Ariel 4070000, \underline{Israel}\\
                 orka@ariel.ac.il

            \and
}
\date{May 10, 2021}

\begin{document}



\maketitle

\begin{abstract}
An independent set in a graph is a set of pairwise non-adjacent vertices. The independence number $\alpha{(G)}$ is the size of a maximum independent set in the graph $G$. The independence polynomial of a graph is the generating function for the sequence of numbers of independent sets of each size. In other words, the $k$-th coefficient of the independence polynomial equals the number of independent sets comprised of $k$ vertices. For instance, the degree of the independence polynomial of the graph $G$ is equal to $\alpha{(G)}$. In 1987, Alavi, Malde, Schwenk, and Erd{\"o}s conjectured that the independence polynomial of a tree is unimodal. In what follows, we provide support to this assertion considering trees with up to $20$ vertices. Moreover, we show that the corresponding independence polynomials are log-concave and, consequently, unimodal. The algorithm computing the independence polynomial of a given tree makes use of a database of non-isomorphic unlabeled trees to prevent repeated computations.
\end{abstract}

\section{Introduction}
\label{sec:Introduction}
The question of whether the independence polynomial of trees is unimodal is still open. Although this is a question that has yet to be answered, it has long been conjectured that  independence polynomials of trees are unimodal 
\cite{mandrescuAndLavitSurvey, alavi1987vertex}.

When discussing the \emph{independence polynomial} of graphs, Alavi et al. \cite{alavi1987vertex} generally showed that the independence sequence of a graph $G$ could be not unimodal. For example, the graph ${K_{25}}+4{K_2}$ has an independence sequence of $\{1, 33, 24, 32, 16\}$, hence, its independence polynomial of $1+33x+24x^2+32x^3+16x^4$ is not unimodal.

Our proof is mainly based on three algorithms. First, an algorithm that generates a unique identifier for a given rooted
tree based on canonization. Second, an algorithm that computes the independence polynomial of a given tree. Third, a main algorithm that iterates over all of the trees with up to 20 vertices and store them in a database along with additional information.

We created a database that contains all of the unlabeled and non-isomorphic trees with up to $20$ vertices. We saved the following details about each tree in our database: a sorted array of degrees, number of vertices, independence polynomial, a unique identifier (the primary key of database) and a Boolean flag representing whether the independence polynomial is unimodal. We use these details to prove our claim and to analyze the research results.

According to Cayley’s formula \cite{caylesFromula}, the amount of labeled trees with $n$ vertices is $n^{n-2}$. Therefore, the number of labeled trees with up to $20$ vertices equals to $\sum_{k=1}^{n} k^{k-2}$. However, in our case, we would like to examine only non-isomorphic unlabeled trees. Hence, Cayley’s formula gives only an upper bound for the number of trees we interested in. For instance, the number of all the non-isomorphic unlabeled trees with up to $20$ vertices is $1,346,025$ \cite{n.j.a.sloane}.
Further, we show that all trees with up to $20$ vertices have unimodal independence polynomials by running a simple SQL query. The query counts the number of trees whose unimodal flag is False. The count returns the result 0, meaning that all of the database trees have unimodal independence polynomials. Also, we ran an SQL query that validates whether all the trees in our database have log-concave independence polynomials.  


\section{Related Work}
\label{sec:related-work}
Logarithmically concave and unimodal sequences take place in many areas of mathematics, such as algebra and combinatorics \cite{[6],[26]}. As a
well-known example, the sequence of binomial coefficients is unimodal. 
As been said at the beginning of the introduction, the conjecture that the independence polynomial of trees is unimodal was never proved. 

However, for other different types of graphs, it was proved that their independence polynomial is unimodal.

Hamidoune showed that the independence polynomial of every graph $G = (V, E)$, where $G$ is claw-free, is log-concave \cite{OnTheNumbersOfIndependentK-SetsClawfree}.
One can easily see that the independence polynomial of a line graph, path graph, cycle graph, barbell graph, and more classes of claw-free graphs is unimodal. 
Moreover, Wang and Zhu showed that the independence polynomial of each vertebrated graph is log-concave, and, hence, unimodal \cite{OntheChina}. 
In \cite{OnUnimodalityOfIndependencePolynomialsOfSomeWellCoveredTrees, OnWellCoveredTreesWithUnimodalIndependencePolynomials}, Levit and Mandrescu proved the unimodality of independence polynomials  for some well-covered trees, including centipedes and well-covered spiders.

Unlike the methods used to prove the above results, in this paper we provide a proof by computation using the brute-force method. We split our statement into a finite number of cases ($1,346,025$ in total \cite{n.j.a.sloane}) and check each case.

\section{Preliminaries}
\label{sec:Preliminaries}
This section provides basic definitions regarding the model explored throughout this paper and an introduction to the methods utilized to prove our statement.
\subsection{Independence Polynomial}
An \textit{independent set} of a graph is a set of pairwise non-adjacent vertices. The independence number $\alpha(G)$ is the size of a maximum independent set in the graph $G = (V, E)$, where $G$ is a finite, connected, undirected, loop-less graph, with vertex set $V$ and edges set $E$. Following Gutman and Harary (1983) \cite{GutmanAndHarary1983}, the \emph{independence polynomial} of $G$ is defined to be:
\begin{equation}
    \label{eq:independent-poly}
    I(G;x) = \sum_{n=0}^{\alpha(G)}{s_n}x^{n}={s_0}+{s_1}x+{s_2}x^{2}+...+{s_{\alpha(G)}}x^{\alpha(G)}
\end{equation} 
where  \({s_k}\) denotes the number of independent sets of cardinality k in graph G, for \(k\in [0, \alpha(G)]\cap N\). Note that ${s_0} = 1$ \cite{OnUnimodalityOfIndependencePolynomialsOfSomeWellCoveredTrees}. Readers can refer to a survey on \emph{independence polynomials} in \cite{mandrescuAndLavitSurvey}.
A finite sequence ${{s_k}_{k=0}^{\alpha(G)}} \subseteq \field{N}$   is said to be: 
\begin{itemize}
    \item unimodal $\iff \exists n \in \{0, 1,...\alpha(G)\}$ such that:     
    ${s_0}\leq{s_1}\leq{s_2}\leq...\leq{s_n}\geq{s_{n+1}}\geq{s_{n+2}}\geq...\geq{s_{\alpha(G)}}$
    \item  log-concave $\iff\ \forall n \in \{1, ... {\alpha(G) -1 }\}\,\ ({s_n})^2\geq{s_{n-1}}{s_{n+1}} $
\end{itemize}

If \(G=(V,E)\) is a tree and \(|V|\subseteq\{1, 2,... 20\}\), the independence polynomial sequence of coefficients \(({{s_k})_{k=0}^{\alpha(G)}}\) is a log-concave (therefore, an unimodal) sequence (see, for instance, \cite{mandrescuAndLavitSurvey}).\\
Readers can refer to \cite{logConcaveIndepenecePolynomial} for reviewing log-concave independence polynomial concerning some graph products.
\\

Throughout the entire paper, $T=(V, E)$ 
is an undirected, finite tree, with indexed vertex set $V=V(T)$ and edges set $E=E(T)$.\\ The operation that will be applied in order to delete any arbitrary set of vertices $V'\subseteq V$ and their edges, is denoted by $T-V'=T'=(V'', E') $, when $V'' = \{v\in{}V : v\notin{}V'\} $ and $E' = \{e=(u, v)\in{}E : v,u\notin{}V'\}$.\\
\\
In order to prove the main result of the paper, we use the two following properties of the \emph{independence polynomials}:\\
\begin{gather}
\begin{gathered}
\label{eq:independent-poly-rec}
 I(G;x)=I(G-v;x)+x \cdot{} I(G-N[v];x) : v\in V(G)\\
\forall{v} \in V(G), N[v] =\{u\in V(G) : (u, v)\in E(G)\} \cup \{v\}
\end{gathered}
\end{gather}

where $N[v]$ is the neighborhood of the vertex $v$ including $v$ itself, and
\begin{gather*}
G-v = (V(G)\setminus\{v\}, E(G)\setminus e(v))\ : \ e(v) = \{(u,v)\in E(G)\} \\
G-N[v] = (V(G)\setminus N[v], E(G)\setminus e(N[v]))\ : \\  e(N[v])=\{(u,v)\in E(G)\}\cup{ \{(u,k)\in E(G-v) \cap (u, v)\in E(G)\}}
\end{gather*}

\begin{gather}
\begin{gathered}
\label{eq:independent-poly-power}
I({G_1}\cup{G_2};x)=I({G_1};x)\cdot I({G_2};x) \ \cite{mandrescuAndLavitSurvey,CilquePolynomial}
\end{gathered}
\end{gather}
An isomorphism of graphs \({G_1}=({V_1}, {E_1}) \ \cap \ {G_2} = ({V_2}, {E_2})\) is a bijection \(\phi :\ {V_1} \to {V_2}\) such that \(\forall u,\ v \in{V_1},\ (u,v)\in{E_1}\iff(\phi(u), \phi(v))\in{E_2}\)

\section{Algorithms}
\label{sec:Algorithms}
\subsection{Unique-ID-Constructor}
In order to store all of the non-isomorphic unlabeled trees in our database without any duplication,  we constructed a unique identifier for each unlabeled tree. The unique identifier enables us to fetch an unlabeled tree from the database with much ease.\\
The unique identifier of each unlabeled tree is a unique primary key in our database. The algorithm based on the tree canonization explained and proved in  \cite{Isomorphism}. Specifically, one can say that our algorithm implements methods mentioned in  \cite{Isomorphism}.

\begin{center}
\begin{table}[H]
\centering
    \caption{ Variables used in Algorithm 1 (Unique-ID-Constructor).}
    \begin{tabular}{ |p{3cm}||p{7cm}|  }
     \hline
     Variable & Definition\\
     \hline
     dist & A method that returns the distance between two input vertices that is defined by the number of edges in the shortest path connecting them.\\ 
     \hline
     D & An array consisting of the distances between each of the tree's vertices and the tree's root.\\
     \hline
     h & The maximum value of D.\\
     \hline
     IdArr & An array consisting of the unique ID of each sub-tree of the given tree and the unique ID of the given tree itself.\\
     \hline
      newID & A temporary array, consists of the IDs of the given vertex's neighbors, that are also farther away from the root, than the given vertex.\\
     \hline
    $\epsilon$ & An empty word.\\
     \hline
      concatenateAll & A list method that concatenates each of the input list's strings to each other ordered from left to right.\\ 
     \hline
    sortBinary & A list method that sorts its elements by their binary values in descending order.\\ 
     \hline
    \end{tabular}
    \label{tab:algo1}
\end{table}
\end{center}
\begin{algorithm}[H]
\caption{Constructed a unique identifier for unlabeled tree}
 \KwIn{Tree $T=(V, E, r)$}
 \KwOut{$T's$ unique identifier}
 $D \leftarrow{} \{dist(r, u), \forall u\in{}V\}$\;
 $h \leftarrow{} max\{d\in{}D\}$\;
 $IdArr \leftarrow{} \{(\epsilon, \epsilon, ... \epsilon) : |IdArr| = |V|\}$\;
 \For{$i \leftarrow{} h$ to $0$}{
 \ForEach{$v\in{}V \ and \ dist(v, r)=i$}{
  \eIf{$i=h$}{
   $IdArr[v.index] \leftarrow{} '10'$\;
   }{
     $newID = \{\}$ \\
    \ForEach{$u\in{}N(v)$}{
      \If{$dist(u, r)=i+1$}{
        $newID \leftarrow newID \cup \{IdArr[u.index]\}$
   }
      }
    $newID \leftarrow newID.sortBinary()$\;
    $IdArr[v.index] \leftarrow{}concatenateAll(newID)$\;
    $IdArr[v.index] \leftarrow{}concatenateAll(\{'1',IdArr[v.index],'0'\})$\;
    }
  }
 }
 \KwRet{$IdArr[r.index]$}\;
 \end{algorithm}
\BlankLine
The algorithm constructs an array of all the distances from the root to each vertex and stores the maximum distance in $h$ (line 1-2). It then initializes IdArr with empty words, an array with the length of $|V|$ (line 3). Afterward, it iterates from the maximum distance (from the tree's root to another vertex) $h$ to 0 (line 4). For each vertex $v$ that its distance equals to $i$, the algorithm is doing one of two things, depending on $i$'s value (line 6). If $i$ equals to $h$, the algorithm inserts $'10'$ into IdArr at the vertex's index (representing a leaf vertex) (lines 6-7). Otherwise, the algorithm iterates over each of $v$'s neighbors, denoted by $u$, and checks whether the distance between $u$ and $r$ equals to $i+1$ (line 10-11). If so, the algorithm adds the the binary value of $u$'s ID to newID array (line 12). At the end of this process, newID array is being sorted descendingly by binary values (line 13).
 Now, the ID of $v$ is the product of concatenation, from left to right, of each element in newID as a string, and then the algorithm adds $'1'$ to the beginning of $v$'s ID and 0 to its end (line 15-16).
 Once the process is done with $i = 0$, the algorithm returns the root's unique identifier (line 16).
 
\begin{figure}[H]
Figure \ref{fig:my_label} constitutes an example for the algorithm, with $|V| = 4$.
    \centering
    \includegraphics[scale=0.5]{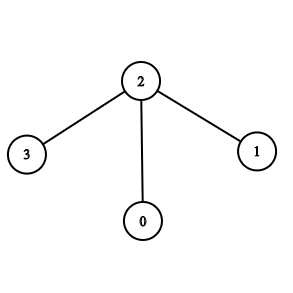}
    \caption{Tree with 4 vertices}
    \label{fig:my_label}
\end{figure}

While vertex '2' is obvious to be the tree's root, and the longest distance from the root is 1 to any vertex in the tree, we want to compute the tree's unique ID. Every single vertex without considering the root itself is the root's neighbor. Thus, at the end of the very first iteration of the main loop in the algorithm, IdArr will become ['10', '10', \(\epsilon\), '10']. At the second iteration, IdArr becomes ['10', '10', '11010100', '10'], as a result of chaining '10' to itself three times, concatenating the output with the existing \(\epsilon\) at the root's index in IdArr, then finalizing the process with adding '1' and '0' to both ends respectively. Meaning, '1101010' is the tree's unique ID.

\subsection{Tree-Independence-Polynomial-Compute}
\begin{figure}[H]
    \centering
    \includegraphics[scale=0.5]{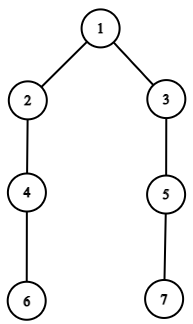}
    \caption{$T=(V, \ E)$}
    \label{fig:computeExample}
\end{figure}

\begin{figure}[H]
    \centering
    \includegraphics[scale=0.5]{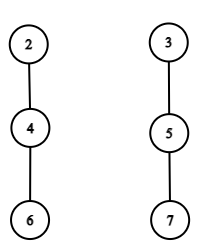}
    \caption{$T-r$}
    \label{fig:computeExample2}
\end{figure}

In this subsection, we will use predefined functions designed to deal with database fetching. We can fetch independence polynomials of trees that have already been computed. Furthermore, as already been said, we can check for every input if an isomorphic tree to it exists in the database.
\begin{center}
\begin{table}[H]
\centering
    \caption{ Variables used in Tree-Independence-Polynomial-Compute.}
    \begin{tabular}{ |p{3cm}||p{7cm}|  }
     \hline
     Variable & Definition\\
     \hline
     DB & The database which consists of the computed trees' independence polynomials and their unique IDs. \\
     \hline
    fetchPolynomial & A database method that returns the matching independence polynomial for the input unique ID (returns NULL if the unique ID doesn't exist in the database). \\
     \hline
     exists & Holds the output of the fetchPolynomial method. \\
     \hline
    $T'$ & An array that consists of every tree component in the output forest from $T-{r}$\\
     \hline
    $T''$ & An array that consists of every tree component in the output forest from $T-N[{r}]$\\
     \hline
     $P_1$ & The product of all trees', in $T'$, independence polynomials.\\ 
     \hline
     $P_2$ & The product of all trees', in $T''$, independence polynomials.\\ 
     \hline
    \end{tabular}
    \label{tab:algo2}
\end{table}
\end{center}
\begin{algorithm}[H]
\caption{Tree-Independence-Polynomial-Compute}
\KwIn{$T=(V,E, r, uid)$}
\KwOut{$I(T; x)$, the tree's independence polynomial.}
$exists \leftarrow{} DB.fetchPolynomial(uid)$\;
\eIf{$exists \neq{} NULL$}{
 return $exists$\; }{
 \If{$|V| = 0$}{return $1$\;}
 \If{$|V| = 1$}{return $1+x$\;}
 $T' \leftarrow{} T - r\; : $ \
 $T' = \{T'_1 \cup{} T'_2 \cup{}...\cup{} T'_n : \forall 1 \leq{} i \leq{} n, T'_i=Tree\}$\;
 $T'' \leftarrow{} T - N[r]\; : $\ 
 $T'' = \{T''_1 \cup{} T''_2 \cup{}...\cup{} T''_m : \forall 1 \leq{} i \leq{} m, T''_i=Tree\}$\;}
 $P_1\leftarrow{1}$\;
 $P_2\leftarrow{1}$\;
 \For{$ 1 \leq{} i \leq{} n$}{
 $P_1 = P_1\cdot{} TreeIndependencePolynomialCompute(T'_i)$\;
 }
  \For{$ 1 \leq{} i \leq{} m$}{
 $P_2 = P_2\cdot{} TreeIndependencePolynomialCompute(T''_i)$\;
 }
\KwRet{$I(T;x)= P_1 + x\cdot{}P_2$}\;
\end{algorithm}
\BlankLine
The algorithm receives a tree. If the tree exists in the database, it fetches the tree's polynomial from the database and returns the output without making any computations (lines 1-3). Otherwise, the algorithm computes the tree's independence polynomial (lines 4-16). If the tree has one vertex or no vertices, the algorithm returns $1$  or $1+x$ respectively, without making any computations (lines 5-8). In any other case, where $|V| > 1$, the algorithm removes $r$ and $N[r]$ from $T=(V,\ E)$ separately and constructs two arrays, consisting of the tree's components.  $T'$ represents the resulted forest of $T-r$ , and   $T''$ represents the resulted forest of $T-N[r]$ (lines 9-10). The algorithm then calls itself recursively with each of the trees in $T'$, computes the output product, and stores it in $P_1$. It then does the same for the trees in $T''$, and stores the output product in $P_2$ (lines 13-16). The algorithm ends with returning the independence polynomial of $T=(V, \ E)$ as $P_1 + x\cdot{}P_2$ (line 17). \\\\
Figure \ref{fig:computeExample} constitutes an example for the algorithm, with $|V| = 7$.
The algorithm receives Figure \ref{fig:computeExample}, and then it constructs $T'$ as an array of two $P_3$ graphs (Figure \ref{fig:computeExample2}), and construct $T''$ as an array of two $P_2$ graphs (Figure \ref{fig:computeExample2}).
\begin{figure}[H]
    \centering
    \includegraphics[scale=0.5]{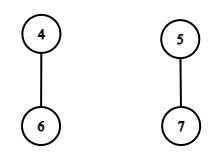}
    \caption{$T-N[r]$}
    \label{fig:computeExample3}
\end{figure}

it then computes the product of all the trees in $T'$ and $T''$ as $P_1$ and $P_2$. This computation happens instantly has $P_2$ graph and $P_3$ graph already exists in the database. Afterwords, the algorithm returns $P_1 + x\cdot{}P_2$ as the independence polynomial of Figure \ref{fig:computeExample}.

\subsection{Main-Algorithm}
We built an algorithm to insert all the non-isomorphic unlabeled trees with up to $20$ vertices into our database.
The algorithm iterates from $2$ to $k \leq 20$, where $k$ stands for the number of vertices in a tree. The first unlabeled tree that has been inserted into the database was $T(V,E) : |V| = 1$, as its independence polynomial $I(T;x) = 1+x$, which is unimodal, and its unique ID is $'10'$. This tree is inserted manually into the database since we are using dynamic programming in the algorithm.\newline
In each iteration, it fetches all of the non-isomorphic unlabeled trees with $k - 1$ vertices from the database, and for each fetched unlabeled tree, it generates $k-1$ trees with $k$ vertices. Each generated tree results from adding a new leaf node to a different node of the tree.
\newline  Then, the Unique-ID-Constructor algorithm (4.1) is being used for each generated tree to enable us to check out the existence of any isomorphic tree, to the iterated one, in the database.
In case there is no isomorphic tree in the database, the Tree-Independence-Polynomial-Compute algorithm (4.2) is being used to compute the iterated tree's independence polynomial $I(T; x)$. \newline
At this point, $I(T; x)$ is being checked for unimodality and whether it is log-concave polynomial. We store all of these in the database and continue to the next tree.
To efficiently insert all the non-isomorphic unlabeled trees with up to $20$ vertices into the database, we used a mechanism to split the iterations on the trees between several instances of the following algorithm. We ran the algorithm non-stop on several Docker containers on a remote server for a week.
\begin{center}
\begin{table}[H]
\centering
    \caption{ Variables used in Algorithm 3 (Main-Algorithm).}
    \begin{tabular}{ |p{3cm}||p{7cm}|  }
     \hline
     Variable & Definition\\
     \hline
     DB & The database which consists of the computed trees' independence polynomials and their unique ID's. \\
     \hline
    fetchAllTrees & A database method that returns an array of the matching trees (re-constructed with unique IDs) for the input number of vertices. \\
     \hline
         isAllTreesInserted & A database method that compares the number \cite{n.j.a.sloane} of inserted trees with the received vertex number to the expected number of trees with that vertex number.  \\
     \hline
     Trees & Holds the output of the fetchAllTrees method. \\
     \hline
    $T$ & A tree from the Trees array.\\
     \hline
    AllTPossible & An array consists of trees, which are made by connecting a new vertex to each of $T$'s vertices, one at a time.\\
     \hline
     $T'$ & A tree from AllTPossible array.\\ 
     \hline
     isTreeExists & A database method that checks if a tree exists in the database.\\ 
     \hline
     P & Holds the output of Tree-Independence-Polynomial-Compute (algorithm 2), which is the input tree's independence polynomial.\\ 
     \hline
     isUni & A polynomial method that returns $True$ if the input polynomial is unimodal, and $False$ otherwise.\\ 
     \hline
     insert & A database method that inserts to the database the new given data.\\ 
     \hline
    \end{tabular}
    \label{tab:algo3}
\end{table}
\end{center}
\begin{algorithm}[H]
\caption{Insert all the non-isomorphic unlabeled trees with up to $20$ vertices}
\KwIn{Array $L$, which consists of the number of non-isomorphic unlabeled trees with up to 20 vertexes.}
\KwOut{None}
\For{$ 2 \leq{} i \leq{} 20$}{
$l \leftarrow{} L[i]$\;
\eIf{$|DB.fetchAllTrees(|V|=i)| = l$}{
 continue\; }{
 $Trees \leftarrow{} DB.fetchAllTrees(|V|=i-1)$\;
\ForEach{$T \in{} Trees$}{
    $AllTPossible \leftarrow{} \{T \cup{\{u\}}_k, \forall 1 \leq{} k \leq{} i-1 \parallel T \cup{\{u\}}_k$ means connecting a vertex to the Tree via the vertex indexed with integer $k$ $\} $\;
    
    \ForEach{$T' \in{} AllTPossible$}{
    \eIf{$DB.isTreeExists(T'.uid)$}{ continue\;}
    {$P \leftarrow$ Tree-Independence-Polynomial-Compute($T'$)\;
    $uni \leftarrow isUni(P)$\; 
    $DB.insert(T'.uid , T'.degreeArr, |V(T')| , P, uni)$\;
    }
   }
   \If{$DB.isAllTreesInserted(|V|=i, l)$}{
 break\; }
  }
 }
}
\end{algorithm}
The algorithm receives an array consisting of the number of non-isomorphic unlabeled trees with $2 \leq |V| \leq 20$. The algorithm iterates from $i=2$ to $20$. If all possible non-isomorphic unlabeled trees with $i$ vertices are already in the database, the algorithm continues (lines 3-4). Otherwise, the algorithm fetches all of the trees with $i-1$ vertices (line 6). It iterates over each tree and creates an array of all the possible trees with $i$ vertices by connecting a new vertex to each existing vertex in the given tree. The algorithm then checks each newly created tree. If the tree does not exist in the database, the algorithm computes its independence polynomial and inserts it with additional information. After iterating over each new tree, the algorithm checks whether all of the possible non-isomorphic unlabeled trees with $i$ vertices are already in the database before continuing to the next tree with $i-1$ vertices (lines 16-17).\\

\begin{figure}[H]
    \centering
    \includegraphics[scale=0.4]{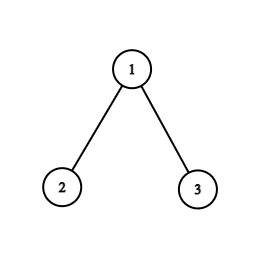}
    \caption{$T=(V, \ E) : |V| = 3$}
    \label{fig:mainExample}
\end{figure}

\begin{figure}[H]
    \centering
    \includegraphics[scale=0.4]{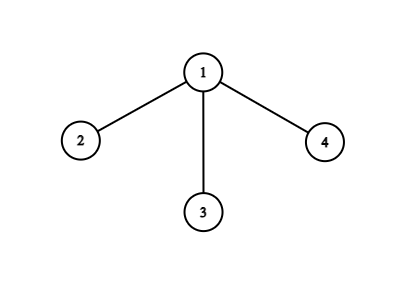}
    \caption{Figure \ref{fig:mainExample} with a new node connected to node number 1}
    \label{fig:mainExample2}
\end{figure}

\begin{figure}[H]
    \centering
    \includegraphics[scale=0.4]{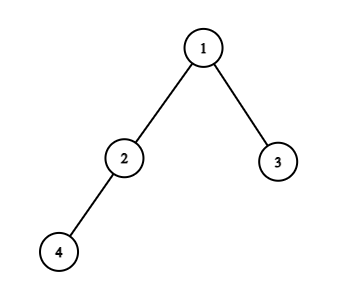}
    \caption{Figure \ref{fig:mainExample2} with a new node connected to node number 2}
    \label{fig:mainExample3}
\end{figure}

\begin{figure}[H]
    \centering
    \includegraphics[scale=0.4]{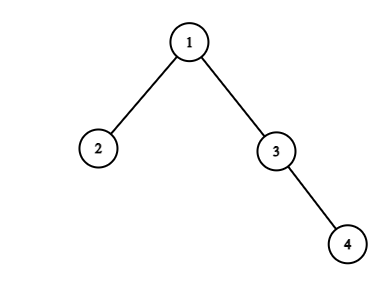}
    \caption{Figure \ref{fig:mainExample2} with a new node connected to node number 3}
    \label{fig:mainExample4}
\end{figure}

Figure \ref{fig:mainExample} constitutes an example for the main algorithm, where $|V| = 3$.
In order to insert all the non-isomorphic unlabeled trees with four vertices, the algorithm fetches all the non-isomorphic unlabeled trees with three vertices from the database (Figure \ref{fig:mainExample}). Then, it iterates on each of the non-isomorphic unlabeled trees with three vertices.
On each tree, the algorithm constructs an array of 3 trees with four vertices by adding a new node to each of the tree's exiting nodes, one at a time (Figures \ref{fig:mainExample2},\ref{fig:mainExample3},\ref{fig:mainExample4}).

The algorithm then iterates on each tree. Starting with Figure \ref{fig:mainExample2}, the algorithm checks if the tree exists in the database. The tree does not exist in the database; therefore, the algorithm computes its independence polynomial and checks if it is unimodal before inserting it with additional data into the database. The same process also happens in Figure {\ref{fig:mainExample3}}  and Figure {\ref{fig:mainExample4}}. However, in Figure {\ref{fig:mainExample4}}, the algorithm continues the iteration because the tree already exists in the database, since (Figure {\ref{fig:mainExample3}} is isomorphic to Figure
{\ref{fig:mainExample4}}).


\section{Proofs}
\label{sec:algorithmsProof}
\subsection{Tree-Independence-Polynomial-Compute}

This algorithm's main requirement is to have in the database it uses all of the trees with $|V|-1$ vertices to compute the independence polynomial of a given tree with $|V|$ Vertices. Thus, we assume that every $T' = (V', E', r', uid') : |V'| < |V|$ exists in the database. 

When the algorithm receives $T=(V,E, r, uid)$, it first tries to fetch its  independence polynomial from the database. If it exists, the algorithm returns the fetched polynomial. Otherwise, the algorithm computes $T$'s independence polynomial as mentioned in (\ref{eq:independent-poly-rec}).
$$T' = T - r = \{{T_1}\cup{T_2}\cup...\cup{T_n} : 1 \leq n < |V|\}$$
$$I(T':x) = I(T-r:x) $$
$$I(T'-r: x) =  I({T_1}\cup{T_2}\cup...\cup{T_n})$$ 
$$I({T'_1}\cup{T'_2}\cup...\cup{T_n}) = I({T_1})\cdot I({T_2}) \cdot ... ({T_n}) = {P_1} \quad (according \ to \ (\ref{eq:independent-poly-power}))$$
$$ \forall{{T_i}} \in \{{T_1}\cup {T_2}\cup...\cup{T_n}\},\quad |{V(T_i)}| < |V(T)| \implies {T_i}\in DB \implies$$
$$I({T_i}:x) = DB.fetchPolynomial({T_i}.uid)$$.

$$T'' = T - N[r] = \{{T_1}\cup{T_2}\cup...\cup{T_n} : 1 \leq n \leq |V| - |N[v]|\}$$
$$I(T'':x) = I(T-N[r]:x)$$
$$I(T'': x) =  I({T_1}\cup{T_2}\cup...\cup{T_n})$$ 
$$I({T_1}\cup{T_2}\cup...\cup{T_n}) = I({T_1})\cdot I({T_2}) \cdot ... ({T_n}) = {P_2} \quad (according \ to \ \ref{eq:independent-poly-power})$$
$$ \forall{{T_i}} \in \{{T_1}\cup {T_2}\cup...\cup{T_n}\},\quad |{V_{T_i}}| < |V_{T}| \implies {T_i}\in DB \implies$$
$$I({T_i}:x) = DB.fetchPolynomial({T_i}.uid)$$

Then it returns ${P_1}+ x \cdot{P_2}$ as $I(T:x)$ as mentioned in \ref{eq:independent-poly-rec};

\subsection{Main-Algorithm}
\textbf{Lemma 5.1} Every tree with $n+1$ vertices, where $n\geq{2}$ can be constructed by adding a new leaf node to an existing node of a tree with $n$ vertices. 

\textbf{Proof.} Suppose a tree $T_i(V_i, E_i)$, where $|V_i|=n+1, n\geq{2}$ can not be constructed by adding a new leaf node to a tree, $T_j(V_j,E_j)$ wheres $|V_j|=n$. Then, removing a leaf node from $T_i$ will not result a tree, $T_j$. This contradicts the fact that $T_j$ is a connected acyclic graph with $|E_j|=|V_j|-1$.


\textbf{Corollary 5.2} A collection of all the trees with $n+1$ vertices can be constructed, wheres $n\geq{2}$, by adding a new leaf node to each existing node of a tree, one at a time, for all the trees with $n\geq{1}$ vertices (\textit{Lemma 5.1}).


\textbf{Theorem 5.3} For every input array $L = \{s_0, s_1, s_2, s_3, ... s_n : s_i = $ the number of non isomorphic trees with $i$ vertices$\}$,  \textit{Main-Algorithm($L$)} computes the independence polynomial of each tree $T=(V,E,uid): 2\leq|V|\leq n$ and store the results in the DB.

\textbf{Proof.} For the sake of this proof, we will use strong induction. \\
\underline{base case:} $n=2:$\\ The main loop in the algorithm will be iterated once. $l = L_2 = 1$. Assuming that the DB is currently holding nothing but the data of $T = (V, E, uid) : |V| = 1$, that has been manually inserted. The first condition in the loop will not be matched, and as a result, the algorithm fetches the only graph existing in the DB, which is, as said, $T = (V, E, uid) : |V| = 1$. Then, the \textit{AllTPossible} array is assigned as the one only possible "extended" tree for a tree with one vertex, that is, the tree with two vertices, and due to the nonexistence of the tree in the DB, the algorithm computes its independence polynomial and insert its findings to the DB. Thus, the statement is true for $n=2$.\\
\underline{Inductive step:} Suppose the statement is true for $n=3, 4, ..., k$.
If $n = k+1$, then, once again, the algorithm skips the first condition that proposed to make the algorithm continue working from the point it last stopped at; for example, in cases of a connection failure with the DB. Afterward, the algorithm fetches every non isomorphic tree $T = (V, E, uid) : |V| = k$ (we suppose all of them exist in the DB) and "extending" each of them to trees with $k+1$ vertices by adding a new leaf node to each existing node of each tree (as Lemma 5.1). The independence polynomial of each extended tree that does not exist in the database is computed before inserting it into the database. After going through each of the trees with $k$ vertices, the algorithm will succeed in computing the independence polynomial of every non isomorphic tree with $k+1$ vertices (Corollary 5.2) and insert the relevant information of each tree into the DB.

\section{Experimental Results}
\label{sec:experimental}
\subsection{Experimental Setup}
\label{sec:WorkingProcess}
We have implemented our algorithms with Typescript and used NodeJS servers to execute our algorithms.\\
We have used PostgreSQL database and a Linux Ubuntu server serving as a docker server for our working environment.\\\\
In order to efficiently insert all of the non-isomorphic unlabeled trees with up to $20$ vertices into the database, we used a mechanism to split the iterations on the trees between several instances of the algorithm. We have split the workload equally, ensuring that each algorithm instance will iterate on different trees from the database. We ran our algorithm 24/7 on several docker containers on our server for a week. Both the database's and the server's specifications are listed in Table \ref{tab:specs}.

\begin{center}
\begin{table}[H]
\centering
    \caption{The database's and the server's specifications.}
    \begin{tabular}{ |p{3cm}||p{3cm}|p{3cm}|  }
     \hline
     & Database & Server\\
     \hline
     CPU   & 8 CPU Core & 1 CPU Core\\
     RAM & 16GB RAM & 2GB RAM \\
     DISK SPACE & 240GB SSD & 50GB SSD \\
     \hline
    \end{tabular}
    \label{tab:specs}
\end{table}
\end{center}

Running several instances of our algorithm working parallel caused the database CPU to reach its limit, thus because our algorithm is frequently fetching data from the database to prevent calculations of trees that already exist in the database. Because of that, we have used the indexing feature of PostgreSQL DB to boost our database performance twice as much.

\subsection{The Unimodality of The Independence Polynomial}
\label{third:WorkingProcess}
As said above, by running a simple SQL query, we showed that all the trees with up to $20$ vertices have a unimodal independence polynomial. The query counted the number of non-isomorphic unlabeled trees whose unimodal flag is False. The count returned the result 0, meaning that all of the trees with up to $20$ vertices have a unimodal independence polynomial. Furthermore, we ran an SQL query that validated that all trees with up to $20$ vertices have a log-concave independence polynomial.

\subsection{Analysis of The Independent Sets}
By applying a simple SQL query on our database, we were able to count, for every cardinally $k$, the number of times it satisfies the following $$\forall T \in DB.Trees, k = ArgMax||S|| : S = \{s_0, s_1, ... s_{\alpha(T)}\} \  and \ I(T;x) = \sum_{i=0}^{\alpha(T)}{s_i}x^{i}$$
With up to $1,346,025$ trees in total \cite{n.j.a.sloane}, we also found that $0 \leq k \leq 11$. Note that we considered the lower cardinality as the maximum, in case of two or more maximums.
\begin{center}
\begin{table}[H]
\centering
    \caption{ The number of trees satisfying the equation above, for each cardinally k:}
    \begin{tabular}{ |p{5cm}||p{5.2cm}|  }
     \hline
     $k$ & Number of trees with most independent sets of cardinally $k$\\
     \hline
     0 & 2 \\ 
     \hline
     1 &  0\\ 
     \hline
     2 & 3 \\
    \hline
     3 & 23 \\ 
     \hline
     4 & 239 \\ 
     \hline
     5 & 3234 \\ 
     \hline
     6 & 58442 \\ 
     \hline
     7 & 851104 \\ 
     \hline
     8 & 420209 \\ 
     \hline
     9 & 12700 \\ 
     \hline
     10 & 68 \\ 
     \hline
     11 & 0 \\ 
     \hline
    \end{tabular}
    \label{tab:cardinalityTable}
\end{table}
\end{center}

\subsection{Different cases and examples}
We have encountered several cases that are worth mentioning. 

\subsubsection{Identical Independence Polynomials}
\textbf{Theorem 7.1} Two trees can have the same $Independence  \ Polynomial$ if they have the same number of vertices. \\
\textbf{Proof.} let $T_1 = (V_1, E_1)$, $T_2 = (V_2, E_2)$ be an arbitrary trees, and \\
$$I_i(T_i:x) = \sum_{n=0}^{\alpha(T_i)}{S_{i_{n}}}x^{n} : i = 1, 2 \ (\ref{eq:independent-poly}) $$ where $S_{1_{1}}$ and $S_{2_{1}}$ denote the number of independent sets of cardinality $1$ in trees $T_1$ and $T_2$ respectively. For every graph $G=(V, E) : |V| = n$, there are $|V|$ independent sets of cardinality $1$, which means $S_{1_{1}} = |V_1|$ and $S_{2_{1}} = |V_2| \ (\cite{K1}) \ \Rightarrow I_1(T_1:x) = I_2(T_2:x) \Leftrightarrow |V_1| = |V_2|. \blacksquare$
\\Also, two trees can have the same independence polynomial, and have or not have the same sorted array of degrees.

\begin{figure}[H]
    \centering
    \includegraphics[scale=0.4]{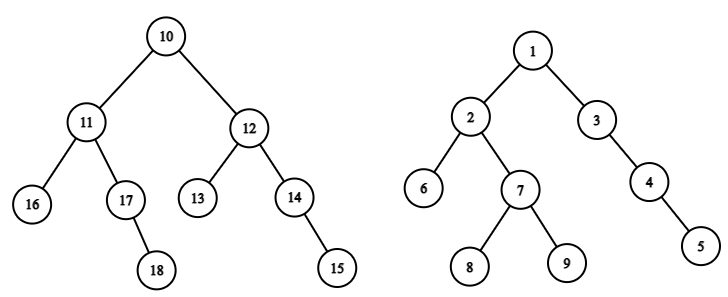}
    \caption{Two non-isomorphic unlabeled trees with the same Independence Polynomial, and the same sorted array of degrees.}
    \label{fig:identicalPolynomialFirst}
\end{figure}
In Figure \ref{fig:identicalPolynomialFirst}, the non-isomorphic trees have the same independence polynomial of $1+9x+28x^2+37x^3+21x^4+4x^5$, and a sorted array of degrees equals to $\{3,3,2,2,2,1,1,1,1\}$.

\begin{figure}[H]
    \centering
    \includegraphics[scale=0.35]{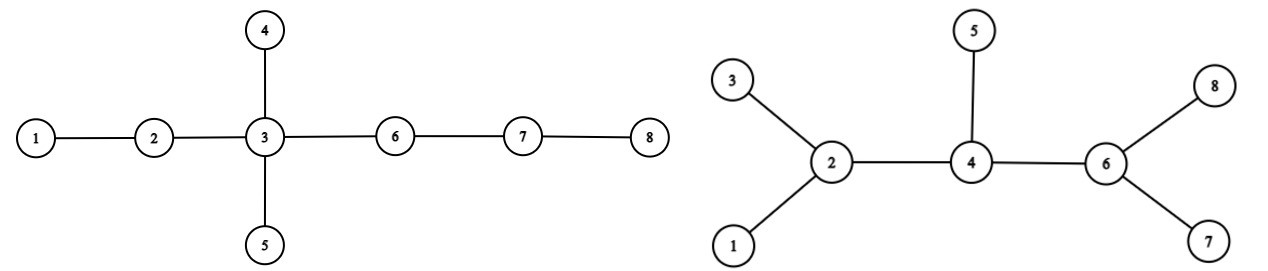}
    \caption{Two non-isomorphic unlabeled trees with the same Independence Polynomial, and different sorted arrays of degrees.}
    \label{fig:identicalPolynomialSecond}
\end{figure}
In Figure \ref{fig:identicalPolynomialSecond}, Both of the trees have an independence polynomial of $1+8x+21x^2+23x^3+11x^4+2x^5$. The left tree posses a sorted array of degrees of $\{4,2,2,2,1,1,1,1\}$, and the right tree posses a sorted array of degrees of $\{3,3,3,1,1,1,1,1\}$. These two trees were presented by Tittmann \cite{GraphPolynomialsTheEternalBook}, and we found them in our database.  

There are three independence polynomials which are most common among the trees with up to $20$ vertices, and each of them can be found 25 times among all the non-isomorphic unlabeled trees with $20$ vertices. 

$$1+20x+171x^2+825x^3+2499x^4+5006x^5+6802x^6+6319x^7+$$ $$3984x^8+1662x^9+435x^{10}+64x^{11}+4x^{12}$$
$$1+20x+171x^2+827x^3+2521x^4+5106x5+7052x^6+6702x^7+$$ $$4360X^8+1900x^9+529x^{10}+85x^{11}+6x^{12}$$
$$1+20x+171x^2+827^3+2523x^4+5125x^5+7127x^6+6863x^7+$$ $$4566x^8+2061x^9+604x^{10}+104x^{11}+8x^{12}$$
\subsubsection{Independence Polynomial with acceding/descending coefficients sequence}
Except for $P_0$ and $P_1$, with the polynomials of $1$ and $1+x$ respectively, we did not find any tree with neither ascending nor descending independence polynomial coefficients sequence.

\subsubsection{Fibonacci sequence of coefficients}
We found four trees of all the non-isomorphic trees with up to $20$ vertices, whose independence polynomials have coefficients that are Fibonacci numbers.
\begin{figure}[H]
    \centering
    \includegraphics[scale=0.4]{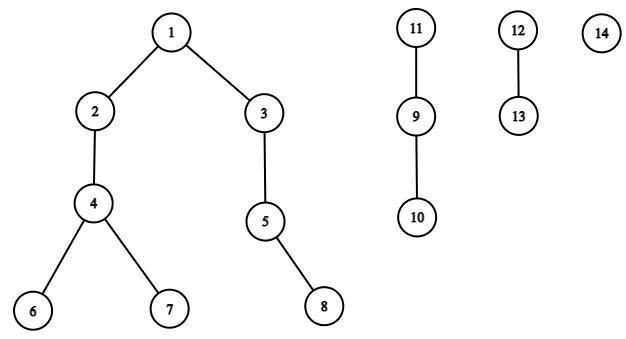}
    \caption{Trees with Fibonacci numbers as independence sequence}
    \label{fig:FibonaccTrees}
\end{figure}

The trees in the above figure have independence Polynomials (from left to right) as follows:
$$1+8x+21x^2+21x^3+8x^4+x^5$$
$$1+3x+x^2$$
$$1+2x$$
$$1+x$$

\subsubsection{Symmetrical Independence Polynomial}
We found that $60$ trees of all the non-isomorphic trees with up to $20$ vertices have an independence Polynomial coefficients sequence that is symmetrical.
 
\begin{figure}[H]
    \centering
    \includegraphics[scale=0.4]{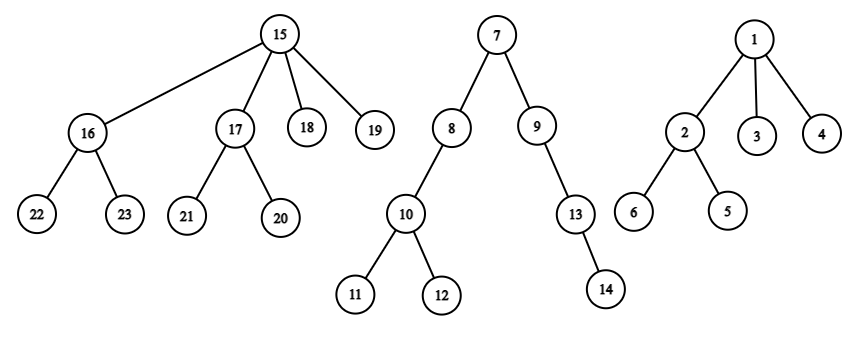}
    \caption{Several trees with symmetric independence polynomial coefficients sequence }
    \label{fig:symmetricalIndependencePolynomial}
\end{figure}
 
The independence polynomials of the trees in the above figure (from left to right) are as follows:
$$1+9x+28x^2+40x^3+28x^4+9x^5+x^6$$
$$1+8x+21x^2+21x^3+8x^4+x^5$$
$$1+6x+10x^2+6x^3+x^4$$

\section{Conclusion and Future Work}
\label{sec:future}
In this paper, we showed that all of the trees with up to $20$ vertices, have log-concave independence polynomials.

We are considering to improve our algorithm so that it will be able to run directly on all of the non-isomorphic unlabeled trees without any duplications. The algorithm will be programmed to generate unique IDs in a sequence that will enable us to efficiently iterating over all of the trees with $n$ vertices. Such a sequence will not include two ids representing the same unlabeled tree.

In addition, we would like to analyze the number of non-isomorphic trees with the same independence polynomials and try to characterize them.



\section*{Acknowledgement}
We would like to greatly thank Professor Eugen Mandrescu from Holon Institute of Technology and Professor Vadim Levit from Ariel University for guiding the research process and the revision process that has led to this paper. This research would not have been possible without their assistance and guidance.

\bibliography{main-V1}

\end{document}